\documentclass[11pt,a4paper]{article}

\usepackage[utf8]{inputenc}

\usepackage{url}
\usepackage[final,dvips]{graphicx}
\usepackage{epsfig}
\usepackage{wrapfig}
\usepackage{float}
\usepackage[hang,small,bf]{caption}
\newcounter{print}
\newcounter{commentbox}
\usepackage[dvips]{graphicx}

\usepackage[scaled=.92]{helvet} 
\usepackage{courier}            

\usepackage{color}

\usepackage{amsmath, amsthm, amssymb}

\def\lsim{\mathrel{\rlap{\lower1.2pt\hbox{\hskip0.6pt${\scriptstyle\sim}$}}\raise2pt\hbox{${\scriptstyle<}$}}}

\newcommand{\bm}[1]{\mbox{\boldmath$#1$}}

\begin{document}


\title{Lepton Mass Effects in Single Pion Production by Neutrinos}

\author{Ch. Berger
\thanks{e-mail: berger@rwth-aachen.de}
\\{\small I. Physikalisches Institut, RWTH Aachen University, Germany} \and
L. M. Sehgal\thanks{e-mail: sehgal@physik.rwth-aachen.de}
\\{\small Institut f{\"u}r Theoretische Physik (E),  RWTH Aachen University, Germany}}
\date{}
\maketitle


\begin{abstract}
We reconsider the Feynman-Kislinger-Ravndal model applied to neutrino-exci\-tation of baryon resonances. The
effects of lepton mass are included, using the formalism of Kuzmin, Lyubushkin and Naumov. In addition we
take account of the pion-pole contribution to the hadronic axial vector current. Application of this new
formalism to the reaction $\nu_\mu + p \to \mu^- + \Delta^{++}$ at $E_\nu \sim$ 1 GeV gives a suppressed
cross section at small angles, in agreement with the screening correction in Adler's forward scattering theorem. Application to the process $\nu_\tau + p \to \tau^- + \Delta^{++}$ at $E_\nu \sim$ 7 GeV leads to
the prediction of right-handed $\tau^-$ polarization for forward-going leptons, in line with a calculation
based on an isobar model. Our formalism represents an improved version of the Rein-Sehgal model,
incorporating lepton mass effects in a manner consistent with PCAC.
\end{abstract}


\section{Introduction}
\label{sec:1}
A new generation of neutrino experiments is under way
that is exploring low energy neutrino reactions such
as $\nu_\mu n \to \mu^- p$ and $\nu_\mu p \to \mu^- p \pi^+$ with unprecedented
statistics and detectors of
high resolution \cite{1}. As an example, the MiniBoone experiment has reported
preliminary results on 40000
events of the type $\nu_\mu p \to \mu^- p \pi^+$ \cite{2}.
These data are providing incisive tests of
theoretical models and revealing features that pose new challenges to theory.
One such feature has been
referred to as a ``deficit of forward muons'' in the reaction
$\nu_\mu + (p,n) \to \mu^- + (p,n) + \pi^+$
and its coherent counterpart $\nu_\mu + {\rm{Nucleus}}
\to \mu^- + \pi^+ + {\rm{Nucleus}}$ \cite{3,4}.
In a recent paper \cite{5} it was pointed out that the inclusion of a non-zero muon mass causes a
suppression of the coherent process in the forward direction, as a consequence of a destructive interference
induced by spin-zero pion exchange.\\

On a different front, lepton mass effects are of obvious importance in reactions
like $\nu_\tau p \to \tau^-  p  \pi^+$
which are being discussed in connection with experiments to detect $\nu_\mu \to
\nu_\tau$ oscillations of atmospheric or laboratory neutrinos \cite{6}.
 The large mass of the $\tau$ lepton
has implications for the angle and momentum distribution of the $\tau^-$ , as well as its polarization. Thus
a formalism that includes non-zero lepton mass is indispensable.\\

In this paper, we propose an extension of the Rein-Sehgal (RS) model \cite{7} for resonance production $\nu_l + N \to l^- + N^*$ that takes account of lepton mass effects. This model, based on the original work of Feynman, Kislinger and Ravndal \cite{8,9}, has been successful in describing data in a variety of neutrino experiments over the past 25 years, and has served as a code for simulating single pion production in the resonance region up to $W \approx 2$ GeV. The attractive feature of the model is its economy: using as input the vector and axial vector form factors of the quasi-elastic channel $\nu_\mu  n \to \mu^-  p$, it provides a unified description of resonance production, embracing nearly 20 resonances in the above mass range. The predictive power of the model derives entirely from SU(6) spin-flavour symmetry and the conservation properties of hadronic weak currents.\\

In introducing lepton mass corrections, we will make use of a formalism developed by Kuzmin, Lyubushkin and Naumov \cite{10}, but will modify it in a significant way that respects the PCAC property of the weak axial vector current.


\section{Brief Review of the RS Formalism $(m_l = 0)$}
\label{sec:2}
We recall the basic features of the RS model in the limit $m_l = 0$ \cite{7,8,9}. The matrix element for a typical resonance excitation process $\nu_\mu + N \to \mu^- + N^*$ is written as
\begin{equation}
  T(\nu_\mu  N \to \mu^- N^*) = \frac{G}{\sqrt{2}} \left[ \overline{u_l} \gamma^\beta (1-\gamma_5) u_\nu \right] \langle N^*|J_\beta^+(0)|N \rangle
  \label{eq:1}
\end{equation}
where $G = G_F \cos \theta_c$, and the hadronic current operator is expressed as
\begin{equation}
  J_\beta^+ = V_\beta - A_\beta = 2 M F_\beta = 2 M \left( F_\beta^V - F_\beta^A \right),
  \label{eq:2}
\end{equation}
$M$ denoting the resonance mass.  The lepton current is expanded in the rest frame of the resonance (RRF) as
\begin{equation}
  \overline{u_l} \gamma^\mu (1-\gamma_5) u_\nu \Big|_\textit{RRF} = -2 \sqrt{2} E_\nu \sqrt{\frac{Q^2}{|\bm{q}|^2}} \left( u e_L^\mu - v e_R^\mu + \sqrt{2uv} e_S^\mu \right)
  \label{eq:3}
\end{equation}
where $Q^2 = - q^2$ is the momentum transfer, and $u$ and $v$ are kinematical factors depending on the initial and final lepton momenta (see Eq.(2.9) of \cite{7}). The vectors $e_L^\mu$, $e_R^\mu$, $e_S^\mu$ may be regarded as polarization vectors of the virtual intermediate boson, corresponding to left-handed, right-handed and scalar polarization,
\begin{equation}
  e_L^\mu = \frac{1}{\sqrt{2}}(0,1,-i,0) ,\ e_R^\mu = \frac{1}{\sqrt{2}}(0,-1,-i,0) ,\ e_S^\mu = \frac{1}{\sqrt{Q^2}}(Q^*,0,0,\nu^*)
  \label{eq:4}
\end{equation}
where $q^\mu = (\nu^*,0,0,Q^* )$ is the momentum transfer 4-vector in the resonance rest frame. The matrix element then takes the form
\begin{equation}
  T(\nu_\mu N \to \mu^- N^*) = -4 G M E_\nu \left\{ \sqrt{\frac{Q^2}{|\bm{q}|^2}} \langle N^* | uF_- - vF_+ | N \rangle + \frac{M_N}{M} \sqrt{2uv} \langle N^* | F_0 | N \rangle \right\}
\label{eq:5}
\end{equation}
with
\begin{equation}
  \begin{split}
    F_+ =& e_R^\mu F_\mu = -\frac{1}{\sqrt{2}}(F_x + iF_y)\\
    F_- =& e_L^\mu F_\mu = \frac{1}{\sqrt{2}}(F_x - iF_y)\\
    F_0 =& \sqrt{\frac{Q^2}{{Q^*}^2}} e_S^\mu F_\mu = F_t + \frac{\nu^*}{Q^*}F_z
  \end{split}
  \label{eq:6}
\end{equation}
(In Eq.\eqref{eq:5}, $M_N$ denotes the nucleon mass, and $\bm{q}$ the 3-momentum transfer in the lab.)
The resulting differential cross section is
\begin{equation}
  \frac{d\sigma}{dQ^2 dW^2} = \frac{G^2}{8 \pi^2 M_N} \kappa \frac{Q^2}{|\bm{q}|^2} \left[ u^2 \sigma_L + v^2 \sigma_R + 2uv \sigma_S \right]
  \label{eq:7}
\end{equation}
with
\begin{equation}
  \begin{split}
    \sigma_{L,R} =& \frac{\pi M}{2 M_N} \frac{1}{\kappa} \sum_{j_z} | \langle N,j_z\mp1 | F_\mp | N^*,j_z \rangle |^2 \delta(W-M)\\
    \sigma_S =& \frac{\pi M_N}{2 M} \frac{1}{\kappa} \frac{|\bm{q}|^2}{Q^2} \sum_{j_z} | \langle N,j_z | F_0 | N^*,j_z \rangle |^2 \delta(W-M),
  \end{split}
  \label{eq:8}
\end{equation}
$\kappa$ denoting the conventional ``flux factor'' $\kappa = (W^2 - M_N^2) / 2 M_N$.
The helicity amplitudes $f_{\pm |2 j_z|} \equiv \langle N, j_z \pm 1 | F_\pm | N^*, j_z \rangle$ and $f_{0 \pm} \equiv \langle N, \pm \frac{1}{2} | F_0 | N^*,\pm \frac{1}{2} \rangle$ are listed in Table II of \cite{7} for all known resonances up to $W = 2$ GeV.
 They are expressible in terms of 3 functions proportional to $G^V(Q^2)$ (these are called $T^V$,
$R^V$ and $S$) and 4 functions proportional to $G^A(Q^2)$ (these are called $T^A$, $R^A$, $B$ and $C$). These 7 functions (called reduced matrix elements or dynamical form factors) are listed in Eq.(3.11) of Ref.\cite{7}.


\section{Lepton Mass Effects due to Nonconservation of Lepton Current}
\label{sec:3}
The RS formalism valid for massless muons has been extended by Kuzmin, Lyubushkin and Naumov (KLN) to the case $m_l \ne 0$ \cite{10}. The first difference is that the lepton in the final state can now have helicity $+$ or $-$, since the lepton current is no longer conserved. So there are six helicity cross sections $\sigma_L^{(\lambda)}$,  $\sigma_R^{(\lambda)}$,  $\sigma_S^{(\lambda)}$, $\lambda = +$ or $-$. In addition, the kinematical coefficients multiplying these cross sections are no longer identical to the factors $u^2$, $v^2$, $2uv$ that occur in the massless case (Eq.\eqref{eq:7}), since the massive lepton satisfies a different energy-momentum relation.
In the notation of KLN, the components of the lepton current in the RRF depend on $\lambda$, and may be written as
\begin{equation}
  \begin{split}
    j_{0(\lambda)}^* =& A_{(\lambda)} \frac{1}{W} \sqrt{1 - \lambda \cos \theta} (M_N - E_l - \lambda P_l)\\
    j_{x(\lambda)}^* =& A_{(\lambda)} \frac{1}{|\bm{q}|} \sqrt{1+\lambda\cos{\theta}} (P_l - \lambda E_\nu)\\
    j_{y(\lambda)}^* =& i \lambda A_{(\lambda)} \sqrt{1+\lambda\cos{\theta}}\\
    j_{z(\lambda)}^* =& A_{(\lambda)} \frac{1}{|\bm{q}| W}\sqrt{1 - \lambda \cos \theta} \left[(E_\nu + \lambda P_l) (M_N - E_l) + P_l (\lambda E_\nu + 2 E_\nu \cos \theta - P_l) \right]
  \end{split}
  \label{eq:9}
\end{equation}
where
\begin{equation}
  A_{(\lambda)} = \sqrt{E_\nu(E_l-\lambda P_l)},
\label{eq:10}
\end{equation}
and the symbols $E_\nu$, $E_l$, $P_l$ and $\theta$ are variables in the lab frame. This modified lepton current can be expanded, as before, in terms of three polarization vectors corresponding to left-handed, right-handed and scalar polarization,
\begin{equation}
  j^{\alpha(\lambda)} = \frac{1}{K} \left[ c_L^{(\lambda)} e_L^\alpha + c_R^{(\lambda)} e_R^\alpha + c_S^{(\lambda)} e_{(\lambda)}^\alpha \right], \quad K = \frac{|\bm{q}|}{E_\nu \sqrt{2Q^2}}
\label{eq:11}
\end{equation}
where
\begin{equation}
  \begin{split}
    e_L^\alpha =& \frac{1}{\sqrt{2}}(0,1,-i,0)\\
    e_R^\alpha =& \frac{1}{\sqrt{2}}(0,-1,-i,0)\\
    e_{(\lambda)}^\alpha =& \frac{1}{\sqrt{Q^2}} \left( Q_{(\lambda)}^*,0,0,\nu_{(\lambda)}^* \right)
  \end{split}
  \label{eq:12}
\end{equation}
Note that the vectors $e_L^\alpha$ and $e_R^\alpha$ are the same as in the case $m_l= 0$. The novelty is in the scalar polarization $e_{(\lambda)}^\alpha$ whose components depend on the lepton helicity $\lambda$. The coefficients $c_L^{(\lambda)}$, $c_R^{(\lambda)}$ and $c_S^{(\lambda)}$ are given by
\begin{equation}
  \begin{split}
    c_L^{(\lambda)} = \frac{K}{\sqrt{2}} &\left( j_x^{*(\lambda)} + i j_y^{*(\lambda)} \right), \quad c_R^{(\lambda)} = \frac{K}{\sqrt{2}} \left( j_x^{*(\lambda)} - i j_y^{*(\lambda)} \right),\\
    &\quad c_S^{(\lambda)} = K \sqrt{\left| \left( j_0^{*(\lambda)} \right)^2 - \left( j_z^{*(\lambda)} \right)^2 \right|}
  \end{split}
  \label{eq:13}
\end{equation}
while the components $Q_{(\lambda)}^*$ and $\nu_{(\lambda)}^*$ are
\begin{equation}
  \begin{split}
    Q_{(\lambda)}^* =& \sqrt{Q^2} \frac{j_0^{*(\lambda)}}{\sqrt{\left| \left( j_0^{*(\lambda)} \right)^2 - \left( j_z^{*(\lambda)} \right)^2 \right|}}\\
    \nu_{(\lambda)}^* =& \sqrt{Q^2} \frac{j_z^{*(\lambda)}}{\sqrt{\left| \left( j_0^{*(\lambda)} \right)^2 - \left( j_z^{*(\lambda)} \right)^2 \right|}}
  \end{split}
  \label{eq:14}
\end{equation}
(Note that starred quantities refer to components in the RRF.) With these definitions, the differential cross section is
\begin{equation}
  \frac{d\sigma}{dQ^2 dW^2} = \frac{G_F^2 \cos^2 \theta_c}{8 \pi^2 M_N}\kappa \frac{Q^2}{|\bm{q}|^2} \sum_{\lambda = +,-} \left[ \left( c_L^{(\lambda)} \right)^2 \sigma_L^{(\lambda)} + \left( c_R^{(\lambda)} \right)^2 \sigma_R^{(\lambda)} + \left( c_S^{(\lambda)} \right)^2 \sigma_S^{(\lambda)} \right]
  \label{eq:15}
\end{equation}
(In the limit $m_l \to 0$, $c_L^{(-)} \to u$, $c_R^{(-)} \to v$, $c_S^{(-)} \to \sqrt{2uv}$, $Q_{(-)}^* \to Q^*$, $\nu_{(-)}^* \to \nu^*$, while $c_{L,R,S}^{(+)} \to 0$.) Remarkably, the helicity cross sections $\sigma_{L,R,S}^{(\lambda)}$ can be calculated exactly as in the RS model
 (using Eq.(2.15) and Table II of Ref.\cite{7}) provided three of the dynamical form factors are modified as follows:
\begin{equation}
  \begin{split}
    S \to S_\textit{KLN} =& \left( \nu_{(\lambda)}^* \nu^* - Q_{(\lambda)}^* |\bm{q}^*| \right)
 \left( 1 + \frac{Q^2}{M_N^2} - \frac{3W}{M_N} \right) \frac{G^V \left( Q^2 \right)}{6 |\bm{q}|^2}\\
    B \to B_\textit{KLN} =& \sqrt{\frac{\Omega}{2}} \left( Q_{(\lambda)}^* 
+ \nu_{(\lambda)}^* \frac{|\bm{q}^*|}{a M_N} \right) \frac{Z G^A \left( Q^2 \right)}{3 W |\bm{q}^*|}\\
    C \to C_\textit{KLN} =& \left[ \left( Q_{(\lambda)}^* |\bm{q}^*| - \nu_{(\lambda)}^* \nu^* \right)
 \left( \frac{1}{3} + \frac{\nu^*}{a M_N} \right) \right.\\
      & \quad \left. + \nu_{(\lambda)}^* \left( \frac{2}{3} W - \frac{Q^2}{a M_N} +
 \frac{n \Omega}{3 a M_N} \right) \right] \frac{Z G^A \left( Q^2 \right)}{2 W |\bm{q}^*|}
  \end{split}
  \label{eq:16}
\end{equation}
Here,
\begin{equation}
  \begin{split}
    \nu^* = E_\nu^* - E_l^* = \frac{M_N \nu - Q^2}{W}, \quad |\bm{q}^*| = \sqrt{Q^2 + {\nu^*}^2},
 \quad a = 1 + \frac{W^2 + Q^2 + M_N^2}{2 M_N W}
  \end{split}
  \label{eq:17}
\end{equation}
As in Ref.\cite{7}, $Z \approx 3/4$ is the renormalization
 factor for the axial vector current, $\Omega = 1.05$ GeV$^2$ is the slope of
the baryon trajectory and $n$ is the number of oscillator quanta in the final resonance. The form factors $G^V \left( Q^2 \right)$ and $G^A \left( Q^2 \right)$ are taken to be dipoles $G^V \left( Q^2 \right) = \left[ 1 + \frac{Q^2}{M_V^2} \right]^{-2}$, $G^A \left( Q^2 \right) = \left[ 1 + \frac{Q^2}{M_A^2} \right]^{-2}$, the default values being $M_V = 0.84$ GeV, $M_A = 0.95$ GeV.


\section{Lepton Mass Effects due to Pion-Pole Term in Hadronic Axial Vector Current}
\label{sec:4}
The KLN formalism described in Sec.\ref{sec:3} takes account of lepton mass corrections stemming from the modification in the components of the lepton current, keeping the hadronic current unchanged. These corrections can be implemented within the RS formalism by redefining the dynamical form factors $S$, $B$ and $C$, as indicated in Eq.\eqref{eq:16}. There is, however, a further effect associated with the nonvanishing lepton mass that remains to be considered. As discussed by Ravndal \cite{9}, the axial hadronic current has, besides the quark current $A_\mu$ defined in Eq.(4.6) of that reference, also a pion-pole contribution, dictated by PCAC, which modifies the axial current as follows:
\begin{equation}
  A_\mu \to \overline{A_\mu} = A_\mu + q_\mu \frac{q^\mu A_\mu}{m_\pi^2 + Q^2}
  \label{eq:18}
\end{equation}
As long as the lepton mass is neglected, the additional term in $A_\mu$ multiplied by the lepton current
gives zero, and so has no effect. If $m_\mu \ne 0$, however, the pion-pole term does contribute. In particular the divergence of $\overline{A_\mu}$ is
\begin{equation}
  q^\mu \overline{A_\mu} = \frac{m_\pi^2}{m_\pi^2 + Q^2} q^\mu A_\mu
  \label{eq:19}
\end{equation}
As shown by Ravndal \cite{9}, this modification of the axial vector current leads to a matrix element for the quasi-elastic process $\nu_\mu + n \to \mu^- + p$ of the form
\begin{equation}
  \begin{split}
    \langle p | \overline{A_\mu} | n \rangle
    = \overline{u_p} \left[ \gamma_\mu \gamma_5 F_1^A (Q^2)
    + q_\mu \gamma_5 F_2^A (Q^2) \right] u_n\\
    \text{with} \quad F_2^A (Q^2) = F_1^A (Q^2) \frac{2M_N + \frac{m_\pi^2}{M_N}}{m_\pi^2 + Q^2}
  \end{split}
  \label{eq:20}
\end{equation}
Neglecting the small correction $m_\pi^2 / M_N$, this result implies an induced pseudoscalar form factor
\begin{equation}
  F_2^A (0) = \frac{2 M_N}{m_\pi^2} g_A(0), \quad g_A (0) = 1.25
  \label{eq:21}
\end{equation}
which agrees with the PCAC result $\left[ F_2^A (0) \right]_\textit{PCAC} =
 \sqrt{2} g_\textit{NN$\pi$} f_\pi / m_\pi^2$ provided 
$g_A(0) 2 m_N = \sqrt{2} g_\textit{NN$\pi$} f_\pi$, which is the Goldberger-Treiman relation ($g_\textit{NN$\pi$}^2 / 4\pi \approx 14$, $f_\pi = 130$ MeV). Thus the inclusion of the pion-pole term in the axial current is mandatory for a satisfactory description of the quasi-elastic matrix element, and will therefore affect the amplitudes for resonance-excitation as well.\\

The effects of the pion-pole term can be incorporated by
 recalculating $e_\mu A^\mu$ and $q_\mu A^\mu$ in the manner described by Ravndal
  (see Eqs.(4.7) and (4.8) of \cite{9}), using the modified axial vector current, Eq.\eqref{eq:18}.
 Only the reduced amplitudes $B(Q^2)$ and $C(Q^2)$ are affected, and the corrected form,
relative to that computed by Kuzmin et al., is
\begin{equation}
  \begin{split}
    B_\textit{BRS}^{(\lambda)} =& B_\textit{KLN}^{(\lambda)} + \frac{Z G_A
\left( Q^2 \right)}{2 W Q^*} \left( Q_{(\lambda)}^* \nu^* - \nu_{(\lambda)}^* Q^* \right)
 \frac{\frac{2}{3} \sqrt{\frac{\Omega}{2}} \left( \nu^* + \frac{{Q^*}^2}{M_N a} \right)}{m_\pi^2 + Q^2}\\
 C_\textit{BRS}^{(\lambda)} =& C_\textit{KLN}^{(\lambda)} + \frac{Z G_A \left( Q^2 \right)}{2 W Q^*} 
\left( Q_{(\lambda)}^* \nu^* - \nu_{(\lambda)}^* Q^* \right) \frac{Q^* \left( \frac{2}{3} W
 -\frac{Q^2}{M_N a} + \frac{n \Omega}{3 M_N a} \right)}{m_\pi^2 + Q^2}
  \end{split}
  \label{eq:22}
\end{equation}
Thus the complete set of dynamical form factors defining the extension of the RS model to take account of muon mass effects is $T^V$, $R^V$, $T^A$, $R^A$ (these are unchanged) together with $S_\textit{KLN}^{(\lambda)}$, $B_\textit{BRS}^{(\lambda)}$ and $C_\textit{BRS}^{(\lambda)}$ given in Eqs.\eqref{eq:16} and \eqref{eq:22}.


\section{Application to $\nu_\mu + p \to \mu^- + \Delta^{++}$}
\label{sec:5}
At energies of order $E_\nu \sim 1$ GeV, the channel $\nu_\mu  p \to \mu^-  p  \pi^+$ is dominated by the $\Delta^{++}$(1232) resonance, and serves as an interesting testing ground for the model discussed in this paper.
\begin{figure}
\centering\epsfig{figure=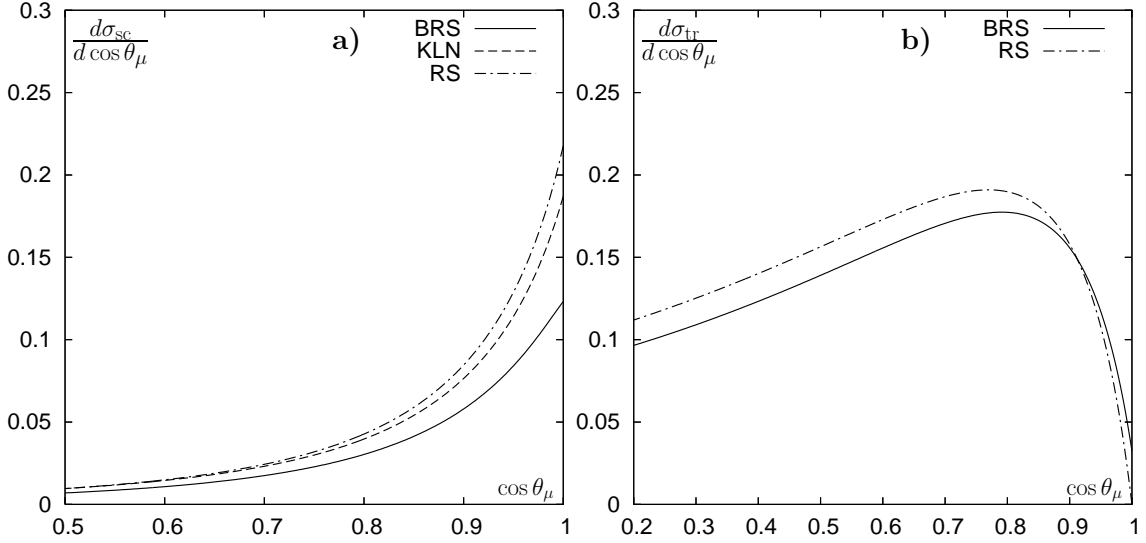,width=\textwidth}
\caption{Differential cross section $d\sigma/d\cos\theta_\mu$ in units of $10^{-38}$ cm$^2$ for the
reaction $\nu_\mu p\rightarrow\mu^-\pi^+p$ at $E=0.7$ GeV. a) scalar cross section for the three models discussed in the text, b) transverse cross section. Here BRS and KLN give identical results.}
\end{figure}

\subsection{Angular Distribution $d\sigma / d \cos \theta_\mu$: Transverse and Scalar Contributions}
To discuss the angular distribution of the muon it is useful to decompose the cross section $d\sigma / dQ^2 dW^2$ into transverse and scalar pieces,
\begin{equation}
  \frac{d\sigma}{dQ^2 dW^2} = \left( \frac{d\sigma}{dQ^2 dW^2}
 \right)_{{\rm tr}} + \left( \frac{d\sigma}{dQ^2 dW^2} \right)_{{\rm sc}}
  \label{eq:23}
\end{equation}
where
\begin{equation}
  \begin{split}
    &\left( \frac{d\sigma}{dQ^2 dW^2} \right)_{{\rm tr}} \sim \sum_\lambda
 \left[ \left( c_L^{(\lambda)} \right)^2 \sigma_L^{(\lambda)}
 + \left( c_R^{(\lambda)} \right)^2 \sigma_R^{(\lambda)} \right],\\
    &\left( \frac{d\sigma}{dQ^2 dW^2} \right)_{{\rm sc}}
 \sim \sum_\lambda \left( c_S^{(\lambda)} \right)^2 \sigma_S^{(\lambda)}
  \end{split}
  \label{eq:24}
\end{equation}
Examination of the kinematical coefficients $c_L^{(\lambda)}$, $c_R^{(\lambda)}$ and $c_S^{(\lambda)}$ given in Eq.\eqref{eq:13} shows that for forward scattering ($\theta_\mu = 0$), the only coefficients that survive, are $c_L^{(+)}$ and $c_S^{(-)}$. The coefficient $c_L^{(+)}$ represents a helicity-flip amplitude, which means, that the only surviving contribution to $\left( d\sigma / d \cos \theta_\mu \right)_{{\rm tr}}$ at $\theta_\mu = 0$ is a helicity-flip cross section which vanishes when $m_\mu \to 0$. By contrast, the scalar contribution $\left( d\sigma / d \cos \theta_\mu \right)_{{\rm sc}}$ at $\theta_\mu = 0$ is determined by $c_S^{(-)}$, and the corresponding helicity-conserving cross section $\sigma_S^{(-)}$, which involves the dynamical form factor $C_\textit{BRS}^{(-)} \left( Q^2 \right)$ given in Eq.\eqref{eq:22}. This form factor is particularly sensitive to the lepton mass because of the pion-pole contribution.\\

The scalar contribution to $d\sigma / d \cos \theta_\mu$ for $E_\nu = 0.7$ GeV is plotted in Fig.1a. The curve labelled `RS' is based on the model of Ref.\cite{7}, with matrix elements calculated for massless muons, but muon mass included in the phase space. The curve `KLN' describes the result of Ref.\cite{10}, based on the dynamical form factor $C_\textit{KLN}^{(\lambda)} \left( Q^2 \right)$, which contains muon mass effects excluding the pion-pole. The full correction is contained in the curve labelled `BRS' based on the dynamical form factor $C_\textit{BRS} \left( Q^2 \right)$ calculated in the present paper. Notice that the scalar cross section at small angles is suppressed by the effects of $m_\mu \ne 0$ (compare `BRS' with `RS').\\
The transverse cross section $\left( d\sigma / d \cos \theta_\mu \right)_{{\rm tr}}$ is plotted in Fig.1b, where we compare the `RS' model with the lepton mass corrected result (here there is no distinction between `KLN' and `BRS'). The difference relative to `RS' reflects the non-conservation of lepton helicity, the cross section containing both $\lambda = +$ and $\lambda = -$. The non-vanishing result in the forward direction represents the helicity-flip component $c_L^{(+)} (\theta = 0) \ne 0$.\\

\begin{figure}
\centering \epsfig{figure=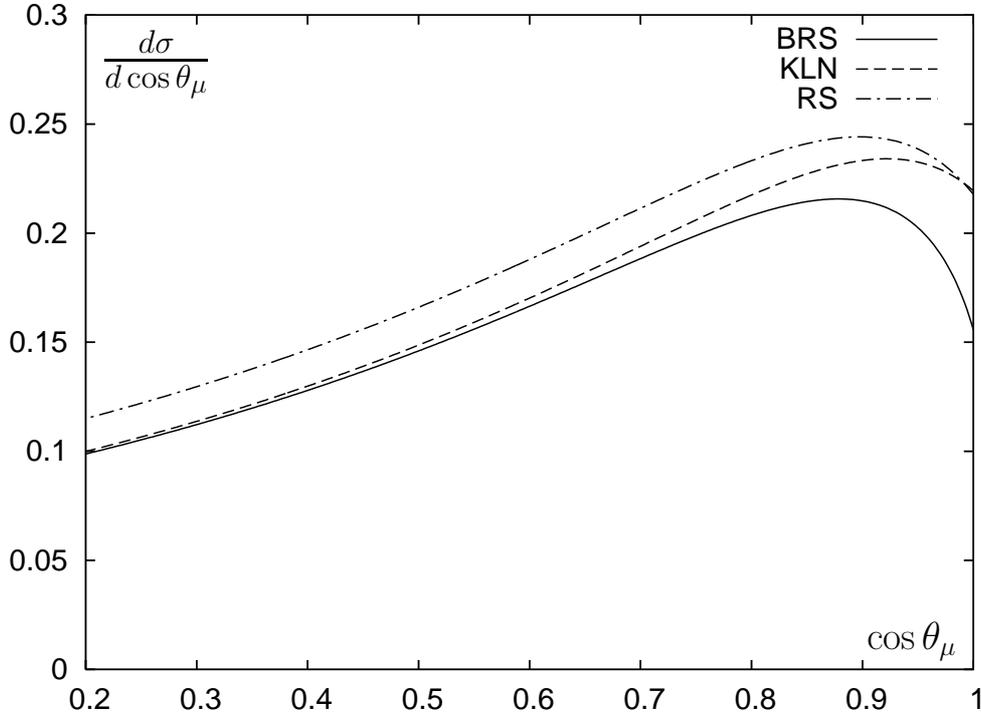,width=0.85\textwidth}
\caption{Differential cross section $d\sigma/d\cos\theta_\mu$ in units of $10^{-38}$ cm$^2$ for the
reaction $\nu_\mu p\rightarrow\mu^-\pi^+p$ at $E=0.7$ GeV resulting from adding the
curves in figures 1a and 1b.}
\end{figure}

Finally, in Fig.2, we add the scalar and transverse pieces to show the full angular distribution $d\sigma / d \cos \theta_\mu$. The muon mass effect is contained in the difference between `BRS' and `RS', and amounts to a suppression in the angular interval $0.9 < \cos \theta_\mu < 1.0$ of $16.5$\%. We have verified that in the limit in which the only mass-corrections retained are those in which a factor $m_\mu^2$ is accompanied by a pole-term $\frac{1}{m_\pi^2 + Q^2}$, the cross section in the forward direction is
\begin{equation}
  \left[ \frac{d\sigma (\theta = 0)}{d \cos \theta_\mu}
 \right]_\textit{BRS} \approx \left( 1 - \frac{1}{2}
\frac{Q_\textit{min}^2}{Q_\textit{min}^2 + m_\pi^2} \right)^2 \left( \frac{d\sigma (\theta = 0)}{d \cos \theta_\mu} \right)_\textit{RS}
  \label{eq:25}
\end{equation}
where $Q_\textit{min}^2 = m_\mu^2 y / (1-y)$, $y = \nu / E_\nu$,
and $\nu^2\gg Q^2_{\rm min}$. We have thus recovered the ``screening'' factor
 $\left( 1 - \frac{1}{2} Q_\textit{min}^2 / (Q_\textit{min}^2 + m_\pi^2) \right)^2$
  which appears in Adler's forward-scattering theorem \cite{11}, and which was invoked
   in Ref.\cite{5} in order to explain the forward muon deficit observed in Ref.\cite{3}.

\begin{figure}
\centering \epsfig{figure=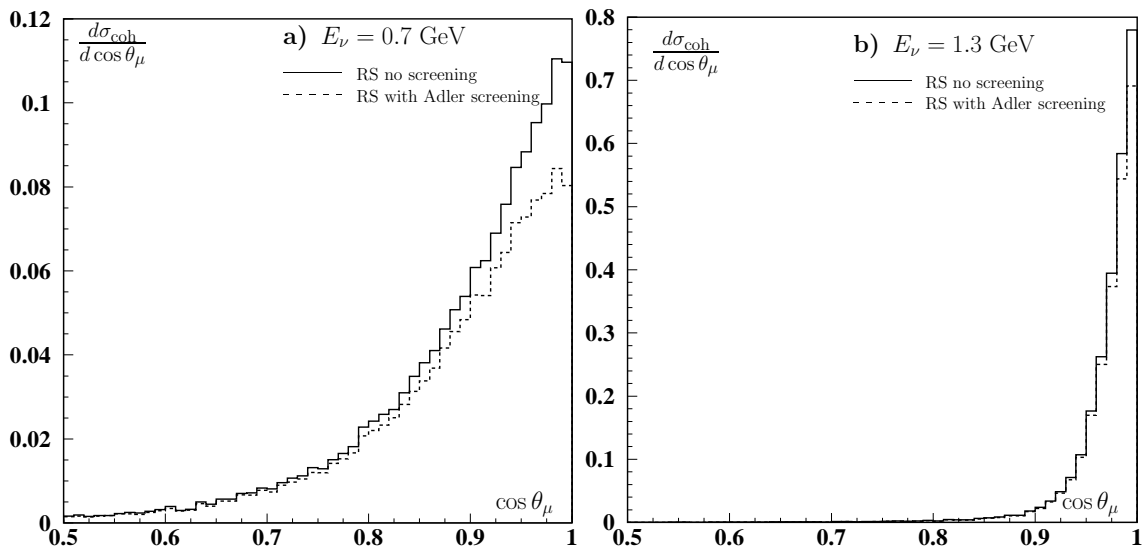,width=0.99\textwidth}
\caption[dum1]{Differential cross section per nucleon
$d\sigma/d\cos\theta_\mu$ in units of $10^{-38}$ cm$^2$ for the
reaction $\nu_\mu+{^{12}C}\rightarrow\mu^-+\pi^++{^{12}}C$. The solid line represents the  coherent Rein-Sehgal model as calculated from (\ref{eq:27}). The dashed line includes
the Adler screening factor (\ref{eq:28}).  a) $E_\nu=0.7$ GeV b) $E_\nu=1.3$ GeV. }
\end{figure}

\subsection{Comparison of Coherent and Incoherent Reactions}
The angular distribution of the $\mu^-$ produced in the incoherent process $\nu_\mu + p \to \mu^- + \Delta^{++} \to \mu^- + \pi^+ + p$ is markedly different from that of muons produced in the coherent process
\begin{equation}
  \nu_\mu + {\rm{Nucleus}} \to \mu^- + \pi^+ + {\rm{Nucleus}}
  \label{eq:26}
\end{equation}
The dynamics of the coherent reaction is governed by the divergence of the weak axial vector current, and the muon angular distribution may be obtained by appealing to Adler's PCAC theorem for small angle scattering. An explicit model for this process was developed in Ref.\cite{12} and leads to the differential cross section (for $m_\mu = 0$)
\begin{multline}
  \frac{d\sigma}{dQ^2 dy dt} = \frac{G^2}{2 \pi^2} f_\pi^2 \frac{1-y}{y} A^2
   \frac{1}{16 \pi} \left[ \sigma_\textit{tot}^{\pi^+ \mathcal{N}} (E_\pi = E_\nu y) \right]^2\\
  \cdot (1 + r^2) \left( \frac{m_A^2}{m_A^2 + Q^2} \right)^2 e^{-b |t|}
  F_\textit{abs} (E_\pi = E_\nu y)
  \label{eq:27}
\end{multline}
where $Q^2$ and $y$ are the momentum and energy transfer variables at the lepton vertex,
 $t$ is the square of the momentum transfer at the nucleus, and the various factors are defined
 in \cite{12}. Using the above formula we have calculated
 $d\sigma / d \cos \theta_l$ for the reaction
 $\nu_\mu + {}^{12}C \to \mu^- + \pi^+ + {}^{12}C$ at $E_\nu = 0.7$ GeV,
 using empirical data on the total and elastic $\pi^\pm p$ cross section to evaluate
 $\sigma_\textit{tot} (\pi^+ \mathcal{N})$ and $F_{\rm abs}$,
 with $\mathcal{N}$ denoting an average nucleon.
 The resulting angular distribution $d\sigma / d \cos \theta_\mu$ is shown in Fig.3a.
 When the effects of muon mass are included, the differential cross section in
 Eq.\eqref{eq:27} is multiplied by the Adler screening factor \cite{5,11}
\begin{equation}
C_\textit{Adler} = \left( 1 - \frac{1}{2} \frac{Q_\textit{min}^2}{Q^2 + m_\pi^2} \right)^2 + \frac{1}{4} y \frac{Q_\textit{min}^2 \left( Q^2 - Q_\textit{min}^2 \right)}{\left( Q^2 + m_\pi^2 \right)^2}
  \label{eq:28}
\end{equation}
This leads to a damping effect in $d\sigma / d \cos \theta_\mu$ at small angles, exhibited in Fig.3a. The corresponding effect at $E_\nu = 1.3$ GeV is shown in Fig.3b. The contrast between the highly peaked distribution in Fig.3a and the corresponding broad distribution of the incoherent process plotted in Fig.2 suggests that the variable $\cos \theta_\mu$ is an appropriate choice for distinguishing between the two processes.

\begin{figure}
\centering \epsfig{figure=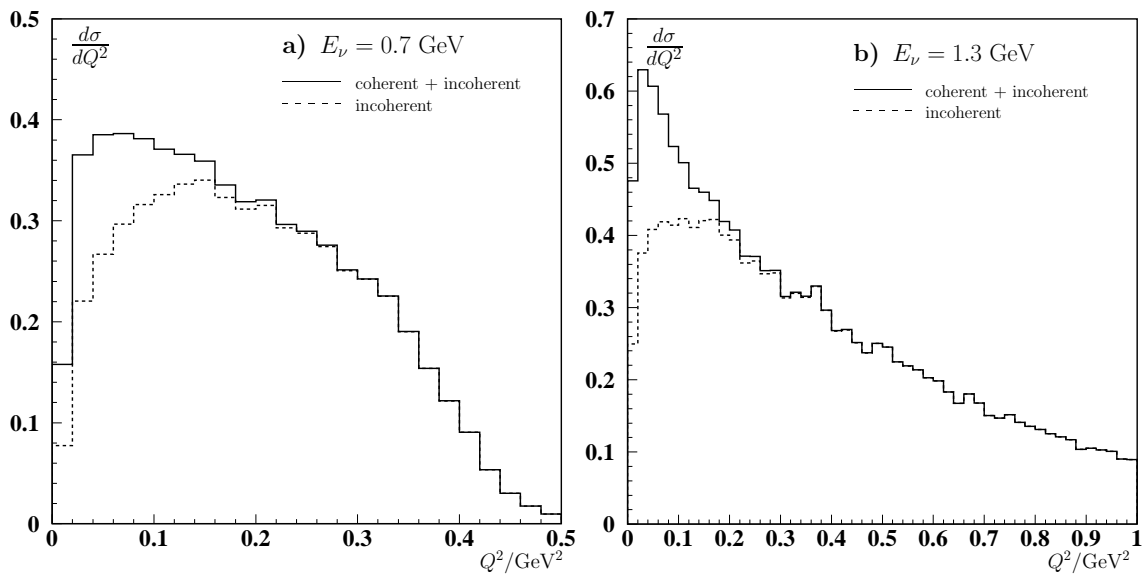,width=0.99\textwidth}
\caption{Differential cross section $d\sigma/dQ^2$ in units of $10^{-38}{\rm cm}^2/$GeV$^2$ for
$\pi^+$  production by  muon neutrinos in charged current reactions. a) $E_\nu=0.7$ GeV  b)
$E_\nu=1.3$ GeV. The incoherent cross section for an average nucleon was calculated in the $\Delta$-approximation, using $\sigma_\textit{tot}^{\pi^+
\mathcal{N}}=(5/9)\sigma_\textit{tot}^{\pi^+p}$.}
\end{figure}

\subsection{Distribution in $Q^2$}
The variable $Q^2$ has also been used as a probe of the form factors involved in the dynamics of $\Delta^{++}$ production, and the distribution in this variable is shown in Figs.4a, 4b for the energies $E_\nu = 0.7$ and $1.3$ GeV, using the BRS model. Also shown is the combined (coherent + incoherent) distribution, where the coherent cross section includes the screening correction, Eq.\eqref{eq:28}. An estimate of the suppression due to muon mass effects may be obtained from Table 1. In the $Q^2$ interval $Q^2 < 0.1$ GeV$^2$, the incoherent cross section for $E_\nu = 0.7$ GeV is suppressed by 14\%. For the coherent process, nearly 80\% of the cross section is in this $Q^2$-bin, and the suppression due to screening is 16\%. Note that this suppression refers to a comparison of the BRS calculation with an unscreened RS model, in which a non-zero muon mass is retained in the
phase space. A stronger suppression is obtained in Ref.\cite{5}, where the screened result is compared with an RS calculation with $m_\mu=0$ everywhere.

It may be added that for neutrino reactions in nuclear targets, a further source of suppression at low $Q^2$ is a possible ``Pauli-blocking'' effect, discussed, for example, in the papers in Ref.\cite{13}. Estimates based on a Fermi gas model indicate that the incoherent process could undergo an additional suppression of $\lsim$ 5\% at $Q^2 < 0.1$ GeV$^2$, as a consequence of such nuclear effects.

\begin{table}
\begin{center}
\begin{tabular}{c|c|c|c|c||c|c|c|c|c}
\multicolumn{5}{c||}{\hspace{7mm}Incoherent scattering}&\multicolumn{5}{c}{\hspace{4mm}Coherent
scattering}\\
&\multicolumn{2}{c|}{$E=0.7$ GeV}&\multicolumn{2}{c||}{$E=1.3$ GeV}
&&\multicolumn{2}{c|}{$E=0.7$ GeV}&\multicolumn{2}{c}{$E=1.3$ GeV}\\
&$\sigma$&$\sigma_{Q^2<0.1}$&$\sigma$&$\sigma_{Q^2<0.1}$
&&$\sigma$&$\sigma_{Q^2<0.1}$&$\sigma$&$\sigma_{Q^2<0.1}$\\
\hline
&&&&&&&&&\\
RS&0.227&0.049&0.504&0.073&RSC&0.173&0.141&0.305&0.242\\
BRS&0.194&0.042&0.483&0.067&RSA&0.147&0.118&0.283&0.222
\end{tabular}
\end{center}

\caption{Integrated cross section for incoherent and coherent neutrino reactions in units
of $10^{-38}$cm$^2$. The index $Q^2<0.1$ indicates integration of
$d\sigma/dQ^2$ between $Q^2=0$ and $Q^2=0.1$ GeV$^2$.
The coherent cross section per \emph{nucleus} has been calculated for $^{12}C$, taking
a Carbon radius of $2.42$ fm.
The line  labelled  `RSC' refers to the coherent Rein-Sehgal model as calculated from (\ref{eq:27}).
The line labelled `RSA' includes the Adler screening factor (\ref{eq:28}).}
\end{table}


\section{Application to $\nu_\tau + p \to \tau^- + \Delta^{++}$}
\label{sec:6}
\begin{figure}[ht]
\centering \epsfig{figure=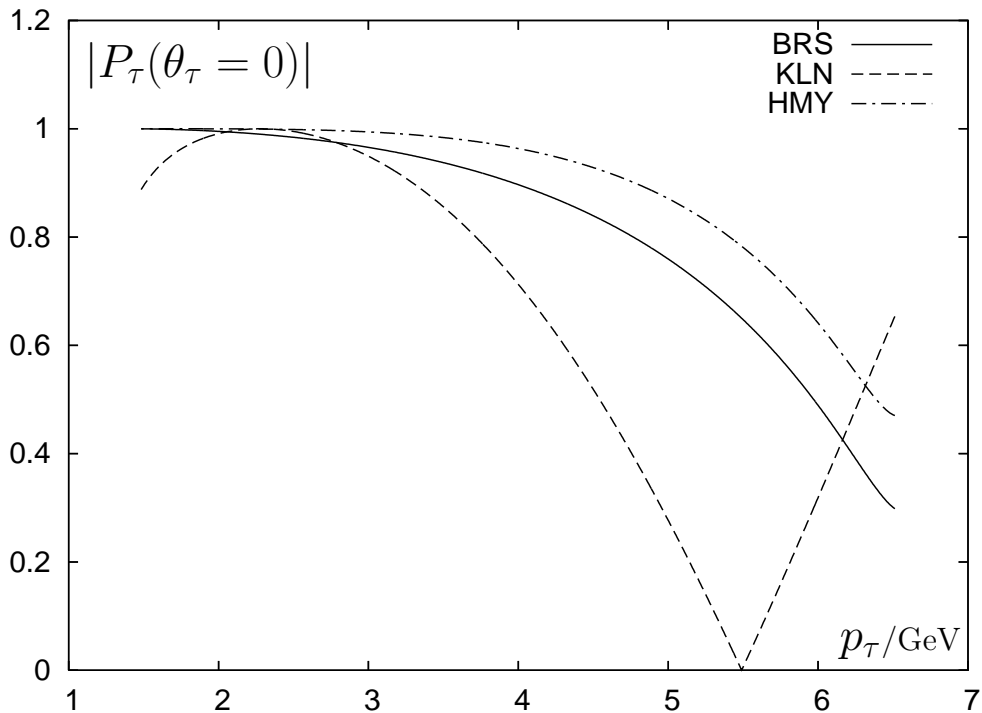,width=0.85\textwidth}
\caption[fig5]{The absolute value of the $\tau^-$ polarization at $0^\circ$ versus
$\tau$ momentum $p_\tau$ for three different
models used in calulating the cross section of $\nu_\tau p\rightarrow \tau^- p \pi^+$. The solid line
(labelled BRS) refers to the model presented in this paper, whereas the dashed line (labelled KLN)
refers to
\cite{10} and the dashed-dotted line (labelled HMY) to \cite{15}. The neutrino energy is $E_\nu=7$ GeV.}
\end{figure}

The consequences of lepton mass corrections are quite dramatic for resonance production by $\nu_\tau$. In particular, it was noted by Kuzmin et al. \cite{14} that in the KLN version of the model, there are significant dynamical effects of $m_\tau \ne 0$, over and above the purely kinematical (phase space) effects. Distributions studied in \cite{14} include $d\sigma / dQ^2$ for $\nu_\tau p \to \tau^- p \pi^+$ for various neutrino energies, and $d\sigma / d p_\tau d \cos \theta_\tau$ at various neutrino energies and angles $\theta_\tau$. One can expect that the `BRS' version of the model will yield predictions that differ from those obtained in \cite{14}. In one instance, Ref.\cite{14} found a result that was in disagreement with a calculation based on an isobar model \cite{15} for the process $\nu_\tau + p \to \tau^- + \Delta^{++}$, which uses a Rarita-Schwinger formalism, and phenomenological form factors constrained by CVC and PCAC. The observable studied was the degree of polarization of the $\tau^-$ in $\nu_\tau + p \to \tau^- + \Delta^{++}$, defined as the fractional difference between helicity $\lambda = +$ and $\lambda = -$ of the final $\tau^-$ lepton. The configuration chosen was $\theta_\tau = 0$ (forward production) and the polarization calculated as a function of $p_\tau$, the $\tau$ momentum, at an energy $E_\nu = 7$ GeV.\\

We have investigated this discrepancy using the present (BRS) version of the RS model which includes lepton mass corrections due to the pion-pole. One expects that the dynamical form factor $C_\textit{BRS}$ would have an impact for forward $\tau^-$ production ($\theta_\tau = 0$), where the screening effect of the pion-pole term should be significant, and the reaction could be dominated by the helicity-flip term $\sigma_L^{(+)}$. In Fig.5, we show our results for the polarization of $\tau^-$, defined as
\begin{equation*}
  P_\tau = \frac{d\sigma (\lambda = +) - d\sigma (\lambda = -)}{d\sigma (\lambda = +)
 + d\sigma (\lambda = -)}
\end{equation*}
as a function of the $\tau$ momentum $p_\tau$, for $\theta_\tau = 0$ and $E_\nu = 7$ GeV.
The curves compare the polarization obtained in the isobar model \cite{15} with the KLN and BRS
 versions of the relativistic quark model.
The BRS result for $P_\tau$ is positive (right-handed helicity),
and compatible with the isobar model, whereas the KLN result
changes sign.
 (The cusp in the KLN curve is due to the fact that Fig.5
  plots the modulus of the polarization.)
We see this as an indication that the incorporation of the pion-pole into the RS model,
 in accordance with the prescription of Ravndal,
  produces lepton mass effects which are in accord with phenomenological approaches based
   on the conservation properties of weak vector and axial vector currents.


\section*{Acknowledgement}
We are indebted to V. Naumov and V. Lyubushkin for providing us with their
 computer routine for the models in Ref.\cite{10} and \cite{15}.
  We also gratefully acknowledge e-mail correspondence regarding numerical problems
  in evaluating the cross section.
\vspace{0.3cm}

\noindent\emph{Note added:} After submission of this paper we saw a
preprint by Graczyk and Sobczyk\cite{16} in which a similar
analysis has been carried out.


\end{document}